\ifdefined\drafting
	\documentclass[aps,prl,notitlepage,superscriptaddress,preprint,floatfix,tightenlines]{revtex4-2}
\else
	\documentclass[aps,prl,notitlepage,superscriptaddress,reprint]{revtex4-2}
\fi

\usepackage[utf8]{inputenc} 
\usepackage{siunitx}
\DeclareSIUnit{\sample}{Sa}
\usepackage{graphicx}
\usepackage{amsmath} 
\usepackage{placeins} 
\usepackage[dvipsnames,svgnames,x11names]{xcolor}
\usepackage{float}
\usepackage{textcomp}
\usepackage{siunitx}
\usepackage{scrlayer-scrpage}
\usepackage{svn-multi} 
\usepackage{amssymb}
\usepackage{amsthm} 
\usepackage{xr} 
\usepackage[colorlinks=true,allcolors=Blue]{hyperref}
\hypersetup{pdfprintscaling=None}

\ifdefined\drafting
	\usepackage[a4paper,left=0.25in,right=1.8in]{geometry}
\else
\fi

\ifdefined\showcomments{}
	\usepackage[commentmarkup=todo,todonotes={textsize=scriptsize,textwidth=1.6in}]{changes}
	\makeatletter \def\@captype{figure} \makeatother 
\else
	\usepackage[final]{changes}
\fi

\newcommand{\markhigh}[1]{\bgroup\markoverwith				
  {\textcolor{#1}{\rule[-.5ex]{2pt}{2.5ex}}}\ULon}

\definechangesauthor[name=Niels Ubbelohde, color=red]{NU}
\definechangesauthor[name=Austris Akmentins, color=ForestGreen]{AA}
\definechangesauthor[name=Slava Kashcheyevs, color=blue]{VK}

\begin{document}


\title{Universal scaling of adiabatic tunneling out of a shallow confinement potential}

\author{Austris Akmentinsh}
\affiliation{Department of Physics,  University of Latvia, 3 Jelgavas street, LV-1004 Riga, Latvia}
\author{David Reifert}
\affiliation{Physikalisch-Technische Bundesanstalt, 38116 Braunschweig, Germany}
\author{Thomas Weimann}
\affiliation{Physikalisch-Technische Bundesanstalt, 38116 Braunschweig, Germany}
\author{Klaus Pierz}
\affiliation{Physikalisch-Technische Bundesanstalt, 38116 Braunschweig, Germany}
\author{Vyacheslavs Kashcheyevs}
\affiliation{Department of Physics,  University of Latvia, 3 Jelgavas street, LV-1004 Riga, Latvia}
\author{Niels Ubbelohde}
\email{Corresponding author: niels.ubbelohde@ptb.de}
\affiliation{Physikalisch-Technische Bundesanstalt, 38116 Braunschweig, Germany}

\begin{abstract}
The ability to tune quantum tunneling is key for achieving selectivity in  manipulation of individual particles in quantum technology applications. In this work we count electron escape events out of a time-dependent confinement potential, realized as a dynamic quantum dot in a GaAs/AlGaAs heterostructure. A universal scaling relation of the escape probability as a function of potential barrier rise time and depth is established and developed as a method to probe tunneling rates over many orders of magnitude reaching the limit of shallow anharmonic confinement. Crossover to thermally activated transport is used to estimate the single time-energy scale of the universal model. In application to metrological single electron sources, in-situ calibrated control signals greatly extend the accessible dynamical range for probing the quantization mechanism. Validation of the cubic potential approximation sets a foundation for microscopic  modeling of quantum tunneling devices in the shallow confinement regime.
\end{abstract}

\maketitle

Tunneling through a potential barrier is a hallmark quantum phenomenon and a fundamental element for evolving quantum technologies, enabling new and limiting existing applications \cite{Chatterjee2021, Pekola2013, Laucht2021}. The exponential sensitivity of tunneling rates to the barrier shape provides discriminative power for quantum state initialization \cite{Kaestner2015,Edlbauer2022}, read out \cite{Hanson2005}, or quantum logic operations \cite{Loss1998}. The full range of tunneling rates that can be exploited by a particular technology is limited by time-energy uncertainty as the minimal barrier height of a confining potential sets the corresponding maximal tunneling rate. This shallow confinement limit constrains the capabilities of tunneling devices for quantum metrology and sensing applications such as the frequency to current conversion \cite{Kaestner2015,Giblin2019} or generation of of ultra short electronic pulses~\cite{Fletcher2012, Bocquillon2014,Bauerle2018}. 
A key to engineering quantum tunneling is quantitative control of the anharmonic confinement potential accessible in situ by experimental realizations. 
While such an achievement proved to be an essential stepping stone~\cite{Martinis1987} for exploiting quantum tunneling at the macroscopic level \cite{Blais2021}, the energy and time scales for fast single-electron control are challenging to probe directly.

Here we map out rates for the last electron escape \cite{MacLean2007} from an emerging quantum dot over many orders of magnitude. Varying the rise time of the confining potential barrier, we employ single electron detection to accurately count electron tunneling events. A scaling relation for the inferred electron capture probability allows to stitch together data spanning several decades of driving speed variation yielding a method to validate the charge capture mechanism over a greatly extended parameter range. Crossover to activated transport is used to estimate the single energy scale which limits the maximal attainable tunneling rate. A minimal microscopic model of ground state tunneling from a cubic potential predicts a universal scaling curve down to the limit of shallow confinement, consistent with our experimental observations.

\begin{figure}[htbp]
	\centering
	\includegraphics{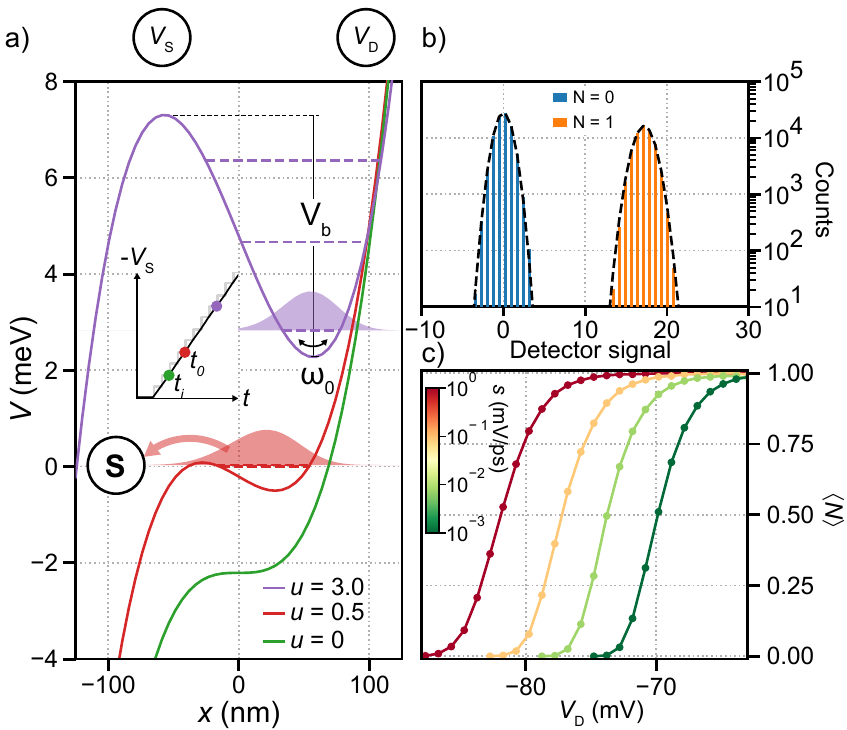}
	\caption{(a) Conceptual sketch of the experiment and the model:   cubic potential $V(x,t)$ for longitudinal confinement  induced by two gate voltages $V_D$ and $V_S (t)$ (linearly ramped, inset), shown here for three values of dimensionless depth $u(t)$, with resonance energies and ground-state probability densities of the corresponding anharmonic oscillator, computed using complex scaling. The zero level is fixed to the Fermi energy of the source by setting $V_0(t)=\mu(t)-E_0(t)+V_b(t)/2$. (b) Example counting histogram of detector signal in units of average detector noise (standard deviation).  (c) Shift of the capture probability $\langle N \rangle$ as a function of $V_D$ due to variation of the ramp rate $s$.  }
	\label{fig:fig1}
\end{figure}

In the experiment, the quantum dot (QD) potential is defined in a GaAs/AlGaAs heterostructure by a shallow etched mesa channel for confinement in the transverse direction, and two metallic top gates inducing two tunneling barriers for longitudinal confinement~\cite{Gerster2018}.
The   corresponding gate voltages, $V_S$ and $V_D$, are tuned such that a very shallow QD can emerge from  the source lead depending on the source gate setting $V_S$, while being isolated from the drain lead at all times. A linear voltage ramp $V_\text{S}(t)=- s\, t$, generated by a filtered digital waveform, is added to the source gate,  while $V_D$ remains constant.  As shown in Fig.~\ref{fig:fig1}a by a sequence of snapshots, the voltage ramp deforms the potential to raise the source tunneling barrier, which then gradually grows and eventually isolates the created QD from the source lead with a decoupling speed proportional to the ramp rate $s$.

To measure the capture probability $\langle N \rangle$ (only $N=0$ and $N=1$ are considered for the number of captured elementary charges $N$) with high accuracy at different ramp rates, a high fidelity counting scheme is employed \cite{Reifert2019}, where the charge state of a large island serving as the source lead is read out by a capacitively coupled detector dot. 
Statistics for $N$ (see Fig.~\ref{fig:fig1}(b)) is obtained by repeatedly executing a sequence of two charge state measurements before and after a single decoupling cycle, followed by a reset operation in which the island is briefly connected to the ground potential. The overall repetition rate of this sequence is \SI{335}{Hz} which ensures a negligible read out error  due to the long integration time independent of the value of $s$. Experiments were performed at a base temperature of \SI{20}{mK} and without a magnetic field applied.
Fig.~\ref{fig:fig1}(c) shows the measured $\langle N \rangle$ for various values of $s$ spanning several orders of magnitude. As a function of $V_\text{D}$, $\langle N \rangle$ transitions from 0 to 1. With increasing decoupling speed, this transition is shifted towards more negative $V_\text{D}$ and therefore towards a shallower potential and more strongly coupled QD, consistent with the logarithmic rise-time dependence \cite{Kaestner2015} observed in a Si-based single-electron ratchet~\cite{Fujiwara2008}.

Our baseline approximation to model $\langle N \rangle$ relies on time-scale separation~\cite{Jauho1994,Fricke2012} between transition rates, which are assumed to follow instantaneously the driving parameters, and the occupation probability $P(t)$ of the QD,  which is eventually taken out of equilibrium with the leads~\cite{Kashcheyevs2010, Kaestner2015}. The corresponding minimal rate equation~\cite{Beenakker1991,Kaestner2008,Fricke2012} is $dP/dt= \Gamma_{\text{in}} [1-P] - \Gamma_{\text{out}} \, P$ where $\Gamma_{\text{in}}(t)$ and $\Gamma_{\text{out}}(t)$ are the charging and the discharging rates for the QD occupation number $N=0 \leftrightarrow 1$. 

The detailed balance relation, $\Gamma_{\text{out}}(t)/\Gamma_{\text{in}}(t) = \exp [\mu(t) / k T_L]$, defines the electrochemical potential difference  $\mu(t)$ between the QD  and the source lead at temperature $T_L$. Large $d\mu/dt$ makes the moment $t_0$ for the onset of backtunneling (lifting of Pauli blockage, $\mu(t_0)=0$) well-defined; the dot switches from $\Gamma_{\text{in}} \gg \Gamma_{\text{out}}$ to $\Gamma_{\text{in}} \ll \Gamma_{\text{out}}$ within $t \in [t_0-\delta t , t_0+\delta t]$ over a timescale $\delta t=k T_L/\dot{\mu}$ much shorter than the timescale $\tau =-(d \ln \Gamma_{\text{out}}/dt)^{-1}$ for the subsequent reduction of $\Gamma_{\text{out}}(t)$ at $t > t_0+\delta t$.
In this limit, the final occupation probability, $\langle N \rangle=P(t\to\infty)$, is determined only by the integral of the escape rate $\Gamma_{\text{out}}(t)$ from $t_0$ till the QD is effectively disconnected, $\Gamma_{\text{out}}(\infty)=0$,

\begin{align}
  \langle N \rangle   & =
 \int\limits_{-\infty}^{\infty} 
    e^{- \int\limits^{\infty}_{t} ({\Gamma_{\text{in}}+\Gamma_{\text{out}}} ) dt' } \frac{d}{dt} \Big(\frac{- \Gamma_{\text{in}}}{\Gamma_{\text{in}}+\Gamma_{\text{out}}} \Big)dt 
    \nonumber \\ &
    \approx
      e^{-\int_{t_0}^{\infty} \Gamma_{\text{out}} \, dt}
      \label{eq:backGen}  
\end{align}
as the fraction under the integral in Eq.~\eqref{eq:backGen} behaves as a delta function $\delta(t-t_0)$ of width  $\delta t \ll \tau$.
Here we assume a fully occupied QD before the escape of the last electron is triggered,  $P(t\to-\infty)=1$. 

\begin{figure*}[htbp]
	\centering
	\includegraphics{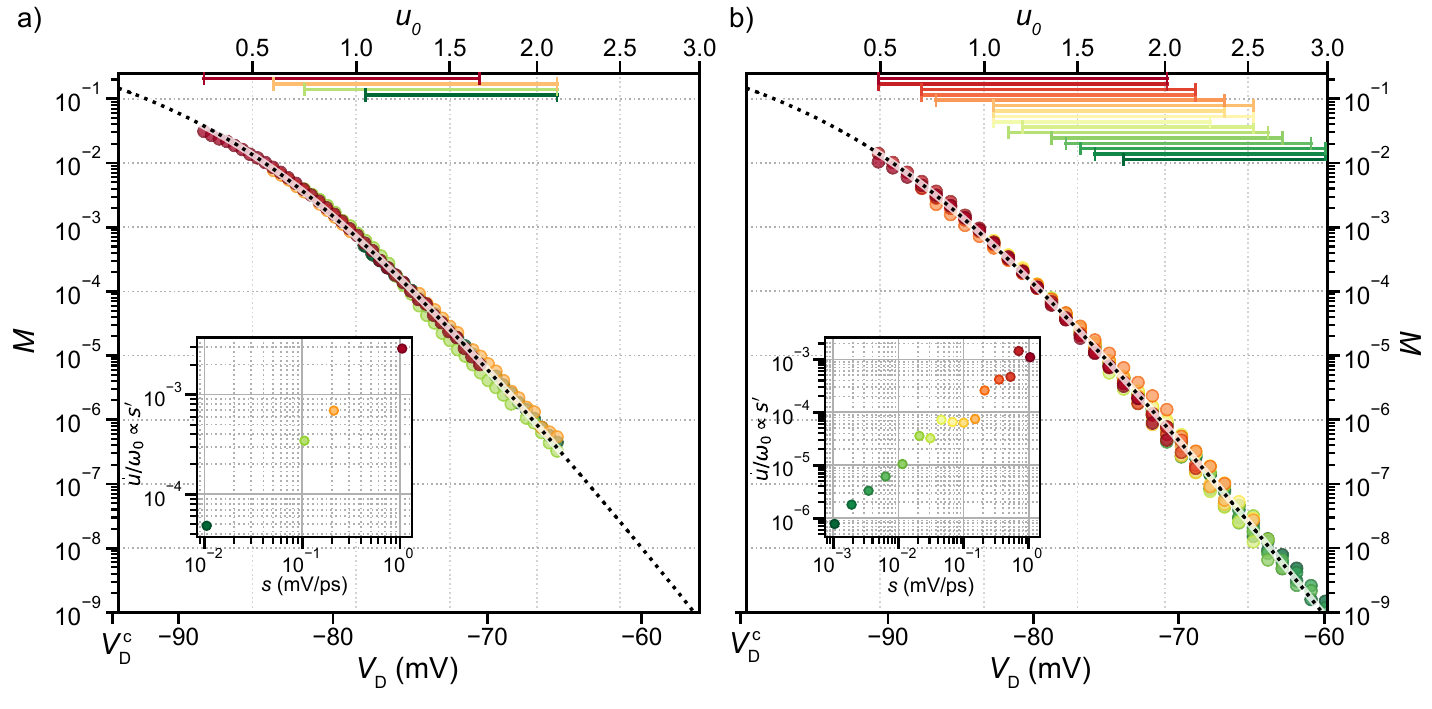}
	\caption{Data collapse of capture probability  plotted as $-(\dot{u}/\omega_0)_i \ln \langle N_i \rangle$ (symbols) as function $V_{\text{D}}$ for filtered  (a) and unfiltered (b) digital voltage ramps ($s$ color mapped). Data points $\langle N_j \rangle$ for $j>1$ appear shifted (multiplied) by  $s'_j/s'_1=(\dot{u}/\omega_0)_j/(\dot{u}/\omega_0)_1$ relative to the reference $\langle N_1 \rangle$. The dotted lines represent a fit to the universal scaling curve $M(u_0)$ given by Eq. \eqref{eq:Generalform} with the inferred $u_0(V_{\text{D}})$ marked on the upper axes. The absolute values of the decoupling speeds $\dot{u}/\omega_0$ as function of the nominal ramp rate $s$ are shown in the inset. The intervals of $u_0/V_D$ measurable for each $s$-parameter are indicated at the upper horizontal axis.}
	\label{fig:fig2}
\end{figure*}

Since the time dependence of  $\Gamma_{\text{out}}$ and $\mu$ is induced by a single parameter~\cite{Kaestner2008}, $V_S(t)$, the integrands in Eq.~\eqref{eq:backGen} always go through the same set of values but possibly with different rates $s'$. The parameter $s'$ is determined by the nominal voltage ramp rate $s$  but accounts for imperfections in signal transmission. Specifically,  for a pair of ramp rates $s_i$ and $s_j$ we expect the corresponding escape rates to be related (up to an irrelevant global shift in $t$) as $\Gamma_{\text{out}}^{(i)}(t)=
\Gamma_{\text{out}}^{(j)}(s_i' \, t/s_j')$ if all other external parameters are kept equal. With this, Eq.~\eqref{eq:backGen} implies a testable scaling relation
\begin{align}  \label{eq:backtunneling}
\left [ \langle N_i \rangle(V_{\text{D}}) \right]^{s'_i} = \left [ \langle N_j \rangle(V_{\text{D}}) \right ]^{s'_j} \, .
\end{align}
In particular, Eq.~\eqref{eq:backtunneling} must hold for any $V_{\text{D}}$ regardless of the functional dependence of $\Gamma_{\text{out}}$ on voltages (which we analyze and model below).  
In order to robustly test Eq.~\eqref{eq:backtunneling} and infer $s'(s)$, we first estimate the exponent matrix $m_{ij}=s'_j/s'_i$  for the predicted power law $\langle N_i \rangle = \langle N_j \rangle^{m_{ij}}$ averaging over data points with different $V_\text{D}$ but common $(s_i, s_j)$. Diagonalizing $m_{ij}$ yields a single dominating eigenvalue with the  corresponding eigenvector proportional to the set of $s'_i$. 

The data collapse of all $\langle N_i \rangle$ for two example data sets (filtered and unfiltered voltage ramps) is confirmed in Fig.~\ref{fig:fig2} by plotting $-\text{ln}\langle N_j \rangle, j>1$, each offset by a factor $ s'_j/s'_1$, on top of $-\text{ln}\langle N_1 \rangle$.
This procedure establishes the relation (insets of Fig.~\ref{fig:fig2}) between the nominal ramp rates $s_i$ and the inferred decoupling speeds $s_i'$ up to overall normalization. 
This result confirms that $\Gamma_{\text{out}}$ follows instantaneously a single parameter which is controlled by the external voltage ramp; the visible discretization steps of the unfiltered waveform in the inset of Fig.~\ref{fig:fig2}(b) underline the sensitivity of the method. 
Establishing this link  to external control voltages and validating the function of key drive parameters despite signal distortions is essential for employing modulated tunneling barriers in high-speed nanoscale devices~\cite{Kaestner2008a, Ahn2017}. 
While at any given parameter setting  of $s$ finite measurement time limits the accurate estimation of $\langle N \rangle$ due to the rarity of events as $\langle N \rangle$ approaches 0 or 1 with $V_D$, the single scaling curve empirically stitched together from measurements at different decoupling speeds now allows to probe the mechanism behind capture over a much larger parameter range. Inferring $\langle N \rangle$ from the backtunneling rate as in Eq.~\eqref{eq:backtunneling} is essential to model the fidelity of charge capture \cite{Kaestner2009, Kashcheyevs2010, Kaestner2015, Fujiwara2008, Giblin2019, Yamahata2021, Hohls2022}, yet so far relied on the linearization of $\ln \Gamma_\text{out}$ over the relevant range of voltages. 
The empirical scaling curve however shows this linearity to be violated at high speeds, i.e.~$\ln (-\ln \langle N \rangle)$ is not linear in $V_{\text{D}}$ in Fig.~\ref{fig:fig2}.

This capping of the exponential growth of the integrated  tunneling rate at increasingly negative $V_\text{D}$ indicates a disappearing source  barrier of an increasingly shallow dot. Confinement emerges near the stationary inflection point of the potential energy for which a cubic approximation is generic~\cite{ Martinis1987,Ankerhold2007,Weiss2021}.
Hence we model the microscopic potential of the QD as $V(x,t)= b\, x^3/3-  F(t) \,x+V_0(t)$, where $x$ is measured from the uniformly moving inflection point in the longitudinal direction and $F(t) \propto  (t-t_i) $ is the lowest order term to capture the transition 
at time $t_i$ marking the formation of a barrier and a well.  The starting shape $V(x,t_0)$ at the onset of backtunneling is set by $F(t_0) \propto (t_0-t_i)$ which is similarly assumed to be linear in the tuning gate voltage $V_{\text{D}}$.
Motion in the transverse direction is assumed to be confined to lowest energy mode and decoupled from $x$.  
Expansion near the inflection point implies non-trivial power laws for the barrier height $V_b=4 F^{3/2} b^{-1/2}/3 \propto (t-t_i)^{3/2}$ and the linear oscillation frequency $\omega_0 =(2/m)^{1/2} (F b)^{1/4}\propto (t-t_i)^{1/4}$ instead of $V_b \propto t$ and $\omega_0=\text{const}$ 
often assumed \cite{Fujiwara2008, Yamahata2019, Yamahata2021} for deeper dots (here $m$ is the effective mass).  We present the results in terms of a dimensionless depth, $u=V_b/(\hbar \omega_0)$, which counts the number of well-localized quasibound states, and the time-independent, device-specific $\Omega_b=\omega_0 \, u^{-1/5}$ which sets the absolute frequency and energy scales. 
The time-independent  speed parameter, $\dot{u}/\omega_0 \propto s'$, sets an absolute scale for the decoupling speed $s'$ for which the relative scale was determined earlier.

In the low-temperature, quantum adiabatic modulation limit we equate $\Gamma_{\text{out}}$ in Eq.~\eqref{eq:backGen} to the decay rate $\Gamma_0$ of the  resonance with the lowest real part $E_n =E_0$ of the corresponding complex energy eigenvalues $E_n - i \hbar \Gamma_n/2$ of the cubic potential~\cite{Caliceti1980}. $\Gamma_0/\omega_0$ is a universal function of $u$ (computed numerically using the complex scaling method \cite{Alvarez1988, Akmentins2022}) 
This results in 
\begin{equation}
\langle N \rangle =\exp \left [- \frac{\omega_0}{\dot{u} } 
\int_{u_0}
\left [ \Gamma_0(u)  /\omega_0(u) \right ]  \, du \right ],
\label{eq:Generalform}
\end{equation} 
where $u_0= u(t_0)$ is the initial depth which is sufficiently well-defined as $\dot{u} \, \delta t  \ll 1$.
The linear relation between $F(t_0)$ and $V_{\text{D}}$ gives the power-law $u_0 = \left[\tilde{\alpha}\left(V_\text{D}-V^c_{\text{D}}\right)\right]^{5/4}$.
Here $V^c_{\text{D}}$, $\tilde{\alpha}$ and $\dot{u}/\omega_0 \propto s'$ constitute fitting parameters mapping the sample-specific function $\langle N \rangle(V_{\text{D}}, s')$ to a parameter-free scaling curve $M(u_0) \equiv -(\dot{u}/\omega_0) \ln \langle N \rangle$ with $\langle N \rangle$ as function of  $u_0$ given by   Eq.~\eqref{eq:Generalform}.
In Fig.~\ref{fig:fig2} a fit of Eq.~\eqref{eq:Generalform} to the empirical scaling  shows excellent agreement with the experiment over the full range of probed ramp rates, including the shallow confinement regime of $u_0<1$, evidence for the anharmonicity of the driven oscillator model. The shape of  $M(u_0)$  contains no device-specific parameters.   Hence  experimental validation  of its universality (Fig.~\ref{fig:fig2} and Fig.~\ref{fig:fig4}b) provides evidence  of the fundamental  microscopic mechanism of electron escape in contrast to the phenomenological decay cascade model \cite{Kashcheyevs2010,Giblin2019}. 
The energy gap protecting this universality is set by the device-specific  scale $\Omega_b$ which we estimate in the following.

For sufficiently fast decoupling speeds, non-adiabatic effects, such as intradot excitation \cite{Kataoka2011, Yamahata2019, Brange2021} and non-Markovian effective temperature  \cite{Flensberg1999,Kashcheyevs2012, Vega2017} are predicted to modify escape dynamics beyond tunneling out of the ground state. 
In our experiment, the quantum adiabaticity is maintained as the increased  decoupling speed shifts the transition $\langle N \rangle  = 0 \leftrightarrow 1$ into the shallow limit ($u_0<1$), hence we use thermal activation  in the regime with several quasibound states ($u_0 >1$) to probe the excitation spectrum and thus estimate the energy scale $\hbar \Omega_b$.

\begin{figure}[htbp]
	\centering
	\includegraphics{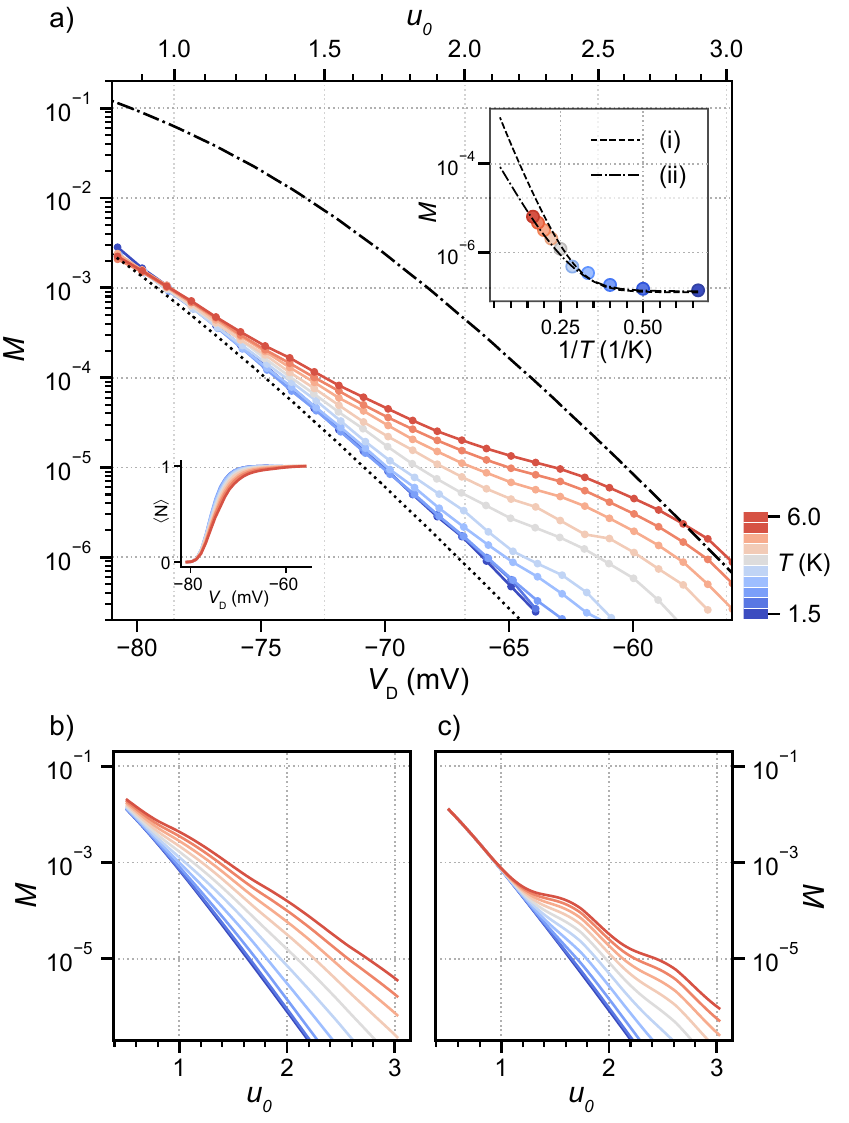}
	\caption{(a) Capture probability as a function of $V_D$, inferred from current measurements, showing thermally activated escape as $T$ is increased up to \SI{6}{K}. The data points are shifted as in Fig.~\ref{fig:fig2}  matching the universal scaling curve (the dotted line). The dashed dotted line shows the model prediction with ground state escape rate replaced by that of the first excited state. Lower left inset: unscaled data, top right inset: Arrhenius plot for $u_0 \approx 2.3$ compared to activated-transport models (i) and (ii) with $\hbar \Omega_b =\SI{1.6}{meV}$ corresponding to $T_0 \simeq \SI{2.9}{K}$. The model predictions as functions of $u_0$ are shown in  (b) and (c) for fast (i) and slow (ii) thermalization limit, respectively.}
	\label{fig:fig3}
\end{figure}

\begin{figure}[htbp]
	\centering
	\includegraphics{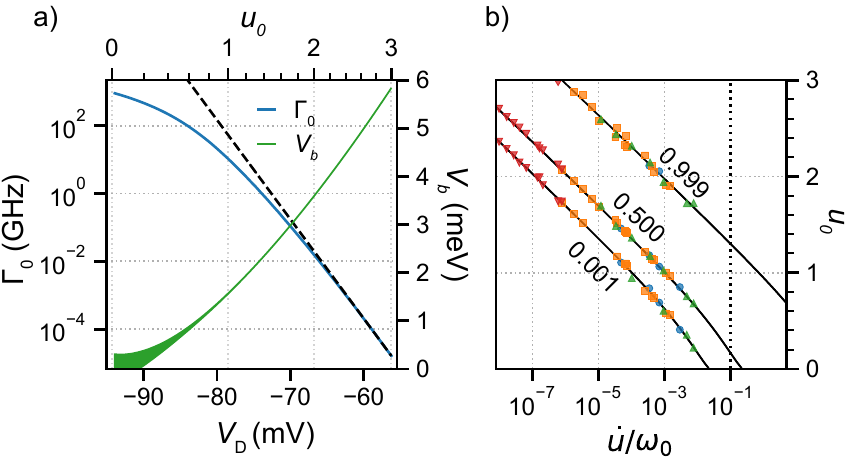}
    \caption{(a) Inferred ground state tunneling rate $\Gamma_0$ and the classical barrier height $V_b$ as a function of gate voltage; width of $V_b$ line shows the quantum uncertainty bracket $\hbar \Gamma_0$. 
    As an indicator of anharmonicity,  exponential extrapolation from the harmonic confinement limit is shown for comparison by the dashed line. (b)  Trade-off between initial depth and speed to reach the same capture probability in the  dimensionless coordinates of the universal model~\eqref{eq:Generalform}. Lines mark three representative levels of $\langle N \rangle$, symbols are various experimental realizations:  a different QD implementation with sinusoidal driving waveform (green $\bigtriangleup$), linear ramp waveform as in Fig.~\ref{fig:fig2}a and b (blue $\bigcirc$ and orange $\square$), and a set of linear ramp waveforms focused on slow decoupling speeds (red $\bigtriangledown$). The vertical line in (b) marks the theoretical limit on adiabaticity ($\dot{u}/\omega_0 < 0.1$, see Ref.~\onlinecite{Akmentins2022}).}
	\label{fig:fig4}
\end{figure}

Thermal broadening of the Coulomb resonances used to read out the charge state and infer the capture probability limits the temperature range accessible to counting to $T<\SI{2}{K}$. Up to this temperature however, no discernible change of the capture probability can be observed. At higher temperatures $\langle N \rangle$ is therefore measured using a precision current amplifier \cite{Drung2015}, detecting the continuous current at a fixed ramp rate of \SI{0.1}{mV/ps} as the captured electrons are emitted towards the drain by further raising the potential \cite{Kaestner2015}. 
Fig.~\ref{fig:fig3} shows $\ln \langle N \rangle$ for temperatures up to \SI{6}{K} shifted by  $s'$ inferred from Fig.~\ref{fig:fig2}. In comparison with the baseline of the counting measurement (black dashed line), the good agreement with the lowest temperatures validates the consistency between the different measurement techniques. Furthermore, a clear crossover temperature $T_0$~\cite{Matveev1996} can be identified on the plateau, $\langle N \rangle \rightarrow 1$, a distinct qualitative difference to behaviour observed in the tail ($V_{\text{D}} <-\SI{75}{mV}$ in the inset). Above this crossover the data points deviate from the universal scaling curve, which conversely corroborates the ground state interpretation of the base-temperature data and contradicts speed-dependent heating.

In order to describe the temperature dependence of the metastability, thermally activated escape via states near the top of the barrier has to be included \cite{Affleck1981,Ankerhold2007,Weiss2021}.
We model this crossover at the level of quantum transition state theory~\cite{Weiss2021} without an explicit model for a heat bath: a Boltzmann distribution with a temperature $T=1/(k_B \beta)$ controls the average over discrete resonances with decay rates $\Gamma_n$ at energies $E_n \lesssim V_b$ and a continuum above the barrier, with decay rate density \cite{Affleck1981} $(2 \pi \hbar)^{-1} \mathcal{T}(E)$ where  $\mathcal{T}(E)= 1/\left\{ 1 + \exp \left[ 2 \pi (V_b-E)/(\hbar \omega_0) \right] \right\}$ is the transmission coefficient in a quadratic approximation. We use the exact solution for the cubic potential to match the semiclassical phase space weights between these two energy ranges as follows: 
\begin{align}
Z \langle \Gamma \rangle &=  \sum_{n=0}^{n_b} \Gamma_n \, a_n  e^{-\beta E_n} + \int_{V_b}^{\infty} \mathcal{T}(E) \, e^{-\beta E} \, \frac{dE}{{2\pi\hbar}} \, , \label{eq:bigGamma}\\
Z &= \sum_{n=0}^{n_b} a_n e^{-\beta E_n} + \int_{V_b}^{\infty} \rho(E) e^{-\beta E} \, dE  \, .  \label{eq:bigZ}
\end{align}
Here the number of  quantum states $n_b+a_{n_b}=
A(u)/(2 \pi \hbar)=18 u/(5 \pi)$ in the phase space area $A(u)$ that is classically confined 
($0 < E < V_b$) is split into the integer ($n_b$) and the fractional ($0\leq a_{n_b}<1$) parts such that $a_n=1$ for $n < n_b$, and $\rho(E)$ is the semiclassical density of QD states above the barrier ($E>V_b$). This model accurately describes crossover from tunneling to hopping~\cite{Affleck1981,Weiss2021} even at $u \sim 1$, see \cite{Akmentins2022} for details.

For the charge capture problem, the thermally activated escape described above needs to be considered in the context of competition between the thermalization timescale and the decoupling time $\tau$. We contrast two opposite extremes of this competition: (i) fast thermalization with respect to decoupling~\cite{Yamahata2021}, where $\Gamma_{\text{out}}(t)$ is replaced by $\langle \Gamma \rangle$ from Eq. \eqref{eq:bigGamma} with time-dependent depth $u(t)$; (ii) slow thermalization, where the number of confined levels $n_b$ and the discrete weights $a_n e^{-\beta E_n}/Z$ are frozen at the initial depth $u=u_0$ and then used in the averaging of $P(\infty)$ over $n$ with $\Gamma_{\text{out}}(t) \to \Gamma_n(t)$. 
Figure~\ref{fig:fig3}(b) and (c) shows the modelling results for both limiting cases, which are compared to the experiment at $u_0 \approx 2.3$ in an Arrhenius plot (inset).
While the intradot population dynamics of the thermally excited states appear to significantly affect the results, both models reproduce the crossover  on the plateau, where the decline of the thermal excitation weight with energy no longer outweighs the competing growth of the tunneling rate. From the experimental data, we deduce $\hbar \Omega_b = \SI{1.6}{meV}$ and hence fix the device-specific microscopic potential (illustrated in Fig.~\ref{fig:fig1}). With $k_B T_0 \simeq 0.16 \, \hbar \Omega_b$~\cite{Akmentins2022} this corresponds to a crossover temperature of $T_0 \simeq \SI{2.9}{K}$. 

In conclusion, matching of the counting data to the universal scaling relation over a broad range of decoupling speeds (Fig.~\ref{fig:fig2}) and the estimation of $\Omega_b$ from the temperature dependence (inset in Fig.~\ref{fig:fig3}) provides an experimental technique for inferring the ground state escape rate $\Gamma_0$ and the barrier height $V_b$ in physical units down to shallow limit where $V_b/\hbar \sim  \Gamma_0 \sim \Omega_b$ and the confinement is eventually lost, see Fig.~\ref{fig:fig4}(a). Using the dimensionless coordinates $u_0$ and $\dot{u}/\omega_0$, different measurements can be put on single universal speed-depth chart, Fig.~\ref{fig:fig4}(b), which follows the level lines of constant $\langle N \rangle$. We find that the devices studied here robustly remain in the regime of ground state tunneling, constrained to operating frequencies of a few GHz by the shallow limit  before non-adiabatic excitations become relevant ~\cite{Akmentins2022}.  
The speed-depth scaling provides a benchmark  to develop electronics manipulating individual particles towards their quantum bandwidth limit. For quantum metrology applications that rely on the escape of excess electrons and capture of the target number of electrons, this work introduces a capability gauge for technology platforms \cite{Kataoka2021, Laucht2021} and a path to certify that the residual error is limited by a universal quantum effect.


\begin{acknowledgments}
Discussions with  Akira Fujiwara, Nathan Johnson,  Peter Silvestrov, and Gento Yamahata are acknowledged. D.R. acknowledges financial support by the Deutsche Forschungsgemeinschaft (DFG, German Research Foundation) within the framework of Germany’s Excellence Strategy-EXC-2123 QuantumFrontiers-390837967. A.A. and V.K are supported by grant no.~lzp-2021/1-0232 from the Latvian Council of Science and the Latvian Quantum Initiative within European Union Recovery and Resilience Facility project no.~2.3.1.1.i.0/1/22/I/CFLA/001. This
work has been supoprted in part by the project 23FUN05 AQuanTEC which has received funding from the European Partnership on Metrology, co-financed from the European Union’s Horizon Europe Research and
Innovation Programme and by the Participating States.
\end{acknowledgments}

\let\oldaddcontentsline\addcontentsline
\renewcommand{\addcontentsline}[3]{}

\bibliographystyle{apsrev4-2}

%

\let\addcontentsline\oldaddcontentsline

\end{document}